\documentclass[aps,prb,superscriptaddress,prb,floatfix,twocolumn,amsmath,amssymb,showpacs]{revtex4-1}
\usepackage{graphicx}
\usepackage{bm}
\usepackage{color}

\begin{document}
\title{Bending mode fluctuations and structural stability of graphene nanoribbons}

\author{P. Scuracchio}
\author{A. Dobry}

\affiliation{
Facultad de Ciencias Exactas Ingenier{\'\i}a y Agrimensura, Universidad Nacional de Rosario and Instituto de
F\'{\i}sica Rosario, Bv. 27 de Febrero 210 bis, 2000 Rosario,
Argentina.}

\date{\today}


\begin{abstract}
We analyze the thermal fluctuations of a narrow graphene nanoribbon.
Using a continuum, membrane-like model we study the height-height correlation functions and the destabilization modes corresponding to two different boundaries conditions: ribbons which are fixed or free on the edges.
For the first situation, the thermal spectrum has a gap and the correlations along the ribbon decay exponentially. Thermal fluctuations produce only local perturbations of the flat situation. However, the long range crystalline order is not distorted. For free edges, the situation changes as thermal excitations are gapless. The low energy spectrum decouples into a bulk and an edge excitation. The bulk excitation tends to destabilize the crystalline order producing an homogeneous rippling. Furthermore, we associate the edge mode with a precluding perturbation leading to scrolled edges, as seen in suspended graphene samples.
\end{abstract}

\pacs{61.48.Gh, 63.22.Rc, 81.07.Bc}


\maketitle

\section{Introduction}
The interplay between lattice deformations and electron dynamics will be an important ingredient to understand and control the electronic properties of graphene future devices.
On one side, an external strain applied to graphene produces a pseudo-magnetic field whose effect has been first theoretically predicted\cite{pseudomgtheo} and then experimentally determined\cite{pseudomgexp}. This could be the starting point of a field called straintronics, namely the control of the electronic properties by applying mechanical strain. On the other hand, the intrinsic corrugation observed since the early experiments in graphene suspended samples affects the electron mobility. Moreover, the fluctuations of this corrugation, called flexural phonons, have been proposed to be the source of the intrinsic limit in the electronic mobility of graphene suspended samples\cite{castroflexural}. Certainly, the control of these corrugations is an important question to address in order to improve the electronic mobility in graphene.

When the dimensionality is reduced, height fluctuations are amplified due to the known tendency to instabilities in low dimensions. We expect that thick ribbons with quasi-one-dimensional geometry will have stronger thermal fluctuations than two dimensional systems. These fluctuations will have important effects on the electronic transport and their mechanism should be  identified in order to control and manage the electronic properties of graphene nanoribbons.

It is the goal of the present paper to study the thermal excitations of graphene nanoribbons. We take as a starting point a continuum model, allowing us to account for the long-wavelenght acoustic phonons. Our focus is to understand how different boundary conditions affect the vibrational modes and how these vibrations perturbate the flatness of the static case. We analyze these points by calculating the out-of-plane flexural phonons and the height-height correlation functions of two different situations: clamped and free edges.

This paper is organized as follows: in Section II we introduce the model by taking a continuum limit of a lattice bending system. We also discuss how the appropriate boundary conditions can be taken into account. In Section III we present a general formalism based in a path integral to obtain the correlation functions. In Section IV and V we obtain the out-of-plane phononic spectrum, the correlation functions, and analyze their consequences. Finally, in Section VI we give our conclusions and perspectives.

\section{The model and the boundary conditions}
\label{sec1}
Single- and few-layer graphene are systems of atomic-scale thickness. As such, a continuum elastic theory of thick plates is not directly applicable. However,  mechanical properties, formation of ripples and  the phonon spectrum as the basis of  the electron-phonon interaction, are well described by the elastic energy form of thick plates. The clue to understand this fact  is that the bending rigidity of graphene does not arise from compressions and dilatations of the continuum medium bounded by free surfaces. Therefore, the bending rigidity parameter cannot be obtained from the elastic parameters of the medium; instead, it is an independent quantity\cite{bendingexp}. It is thought that the bending rigidity in graphene is due to the bond-angle and bond-order term associated with the dihedral angles of the underlying C-C interactions \cite{bondieadral}.

This distinction has special significance in the presence of edges, as is the case of the ribbons we are interested on in this work. To make the discussion concrete,
we start from a simplified lattice bending model which has been introduced in the studies of membranes\cite{Lubensky}. The model Hamiltonian is:

\begin{eqnarray}
E_{bend}=-\bar{\kappa}\sum_{<i,j>} (\textbf{n}_i . \textbf{n}_j-1),
\label{bendlatt}
\end{eqnarray}

where $\textbf{n}_i$ is the unit normal vector at the $i$th site of the lattice and $j$ is its nearest neighbor. We use $\bar{\kappa}$ as the bending rigidity parameter in the lattice model.  For small deviations from the flat configuration, we can parametrize the vector position of a point on the surface as:

\begin{eqnarray}
\textbf{r}(x,y)=(x,y,h(x,y)).
\label{rxyh}
\end{eqnarray}

This is called a Monge representation. $h(x,y)$ is the height variable
and the normal unit vector becomes:

\begin{equation}
\textbf{n}(x,y)=\frac{{\textbf{e}}_z-\nabla h}{\sqrt{1+(\nabla h)^2}}=\frac{(-\partial_x h,-\partial_y h, 1)}{\sqrt{1+(\nabla h)^2}},
\label{n}
\end{equation}

where ${\textbf{e}}_z$ is the unit vector in the $z$ direction. To obtain a continuum limit of Hamiltonian (\ref{bendlatt}) we undertake a gradient expansion. We obtain:

\begin{eqnarray}
H_{bend}&=&  \displaystyle\lim_{continuum} \bar{\kappa}\sum_{<i,j>}(1- \textbf{n}_i . \textbf{n}_j)\nonumber\\
&=& \displaystyle\lim \, \frac12  \bar{\kappa}\sum_{<i,j>} (\textbf{n}_i-\textbf{n}_j)^2=\frac{\kappa}{2}\int d^2x \left|\nabla\textbf{n}\right|^2\nonumber\\
&=&\frac{\kappa}{2} \int d^2x \left[(\partial_x^2h)^2+(\partial_y^2h)^2+2 (\partial_x\partial_yh)^2\right],
\label{Hbendcont}
\end{eqnarray}

where we have neglected terms of order $\mathcal{O}(h^2)$ such as $(\nabla^2h)^2$ in the denominator of Eq.\ (\ref{n}) and the one in the integration measure. Explicit geometric contributions of Eq.\ (\ref{Hbendcont}) can be seen if we write it down in the following way:

\begin{eqnarray}
H_{bend}&=&\frac{\kappa}{2} \int d^2x \left[(\nabla h)^2-2 Det(\partial_i\partial_jh)\right].
\label{Hbendcurv}
\end{eqnarray}

The first term is proportional to the square of the mean curvature and the last to the Gaussian curvature, both written in the lowest order of a gradient expansion in $h(x,y)$. In terms of these curvatures, Eq.\ (\ref{Hbendcurv}) is known as the Helfrich form  of the bending energy of a liquid membrane\cite{Helfrich}. The Gaussian curvature is a total derivative term and was neglected in previous studies of the stability of graphene membranes. However, it plays an important role in the search for the appropriate boundary conditions of our ribbon geometry.

If, instead of Eq.\ (\ref{bendlatt}), we had begun with the elasticity problem of a thin plate, the bending energy would have assumed the same expression as Eq.\ (\ref{Hbendcurv})\cite{Landau}. However, in this last case, the bending rigidity would not have been an independent parameter but a function of the elastic modulus of the plate as $\kappa=\frac{Yl^3}{12(1-\sigma^2)}$, being $l$ its thickness, $Y$ the Young's modulus and $\sigma$ the Poisson ratio. Also, the Gaussian curvature of the second term of Eq.\ (\ref{Hbendcurv}) would have been multiplied by $(1-\sigma)$. Regarding the discussion at the beginning of the section, we see that there is a formal connection between the theory of thin elastic plates and the one of a two dimensional membranes. The last can be obtained from the former by taking the bending rigidity as an independent parameter and by setting $\sigma=0$.

In addition to the surface term corresponding to the Gaussian curvature, there is another total derivative term coming from the first part of Eq.\ (\ref{Hbendcurv}), which can be written as:

\begin{eqnarray}
H_{bend}&=& \frac{\kappa}{2}\int d^2x  \Bigg[h \left(\partial_x^4+\partial_y^4+2\partial_x^2 \partial_y^2\right)h+\nonumber\\
&&\partial_x\left(\partial_x h \partial_x^2 h-h\partial_x^3 h-2 h \partial_y^2 \partial_x h\right)+\nonumber\\
&&\partial_y\left(\partial_yh \partial_y^2 h-h\partial_y^3 h+2 \partial_x h \partial_y \partial_x h\right)\Bigg].
\label{Hbendsepsurf}
\end{eqnarray}

Until now we have not specified neither the integration domain nor the physically boundary condition for our problem. We consider a long and narrow ribbon of width $W$ and length $L$ running along the $y$ direction. The domain of integration is therefore given by $-W/2\leqslant x\leqslant W/2$ and $-L/2\leqslant y\leqslant L/2$ with $W\ll L$. Periodic boundary conditions are taken in the $y$ direction. Therefore, the surface term corresponding to the last line of Eq.\ (\ref{Hbendsepsurf}) vanishes. We finally obtain the following bending energy for the ribbon:

\begin{eqnarray}
H_{bend}&=& \frac{\kappa}{2}\int_{-\frac{L}{2}}^{\frac{L}{2}}dy\int_{-\frac{W}{2}}^{\frac{W}{2}}dx
 \left[h \left(\partial_x^4+\partial_y^4+2\partial_x^2 \partial_y^2\right)h\right]+\nonumber\\
&&\int_{-\frac{L}{2}}^{\frac{L}{2}}dy\bigg\{\partial_x h(x,y) [\partial_x^2 h(x,y)]-\nonumber\\
&&h(x,y)\left[\partial_x^3 h(x,y)+2 \partial_y^2 \partial_x h(x,y)\right]\bigg\}_{x=\pm W/2}.
\label{Hbendribbon}
\end{eqnarray}

The last symbol means that the term into braces is the difference between this expression evaluated at $x=\frac{W}{2}$ and at $x=-\frac{W}{2}$. Therefore, the integral runs along the edges of the ribbon. A cancellation of this term could take place in two different situations.  Consider a ribbon with its edges clamped along the $y$ direction. In this case we have:

\begin{eqnarray}
 h(x=\pm\frac{W}{2},y)&=&0\nonumber\\
\partial_xh(x=\pm\frac{W}{2},y)&=&0.
\label{fixbound}
\end{eqnarray}

 The terms multiplying $h(x=\pm\frac{W}{2},y)$ and $\partial_xh(x=\pm\frac{W}{2},y)$ can be interpreted as the force and the torque on the edge of the ribbon\cite{Landau}. Setting these terms to zero means having free edges, and the boundary conditions are then:
 
\begin{eqnarray}
\left(\partial_x^3 +2 \partial_y^2 \partial_x \right)h(x=\pm\frac{W}{2},y)&=&0\nonumber\\
\partial_x^2 h(x=\pm\frac{W}{2},y)&=&0.
\label{freeboundsig0}
\end{eqnarray}

Note that those are the situations (clamped and free) considered in Ref.\ \onlinecite{grapheneacoustic} to obtain the phonon dispersion relations of a nanoribbon. In this reference, the theory of an elastic thin plate is used and  the following boundary conditions are considered for free edges:

\begin{eqnarray}
\left(\partial_x^3 +(2-\sigma) \partial_y^2 \partial_x \right)h(x=\pm\frac{W}{2},y)&=&0\nonumber\\
\left(\partial_x^2+\sigma \partial_y^2\right) h(x=\pm\frac{W}{2},y)&=&0.
\label{freeboundsigne0}
\end{eqnarray}

For $\sigma=0$, Eq.\ (\ref{freeboundsigne0}) is the same as Eq.\ (\ref{freeboundsig0}) as we already remarked. So far we have not discussed about the elastic stretching energy of the ribbon. This is because, in the harmonic approximation, the in-plane modes that correspond to the stretching energy are decoupled from the out-of-plane ones, allowing us to study these situations separately.
 In the next section we will compare the "height-height" correlation function and the mean square of the height corresponding to the two different boundary conditions of Eqs.\ (\ref{fixbound}) and (\ref{freeboundsig0}).

\section{General formalism for correlation functions}
\label{GFCF}
For both clamped and free boundary conditions, the surface term in Eq.(\ref{Hbendsepsurf}) vanishes and the partition function of the system can be written as a path integral of the form:

\begin{eqnarray}
Z&=& \int  \mathcal{D} h\,  e^{-\frac{\kappa}{2kT} \int_{-\frac{L}{2}}^{\frac{L}{2}}dy\int_{-\frac{W}{2}}^{\frac{W}{2}}dx
 \left[h \underbrace{\left(\partial_x^4+\partial_y^4+2\partial_x^2 \partial_y^2\right)}_\mathcal{O}h\right]},\nonumber\\
\label{partfunc}
\end{eqnarray}

integrating over all the paths that fulfill the boundary condition (\ref{fixbound}) or (\ref{freeboundsig0}). It is convenient to
expand the path in the basis of the eigenfunctions of the operator $\mathcal{O}$. Due to the periodic boundary condition in the $y$ direction,we can separate its $y$ dependence. The eigenfunctions assume the form:  $f_{m}^n (x) e^{i q_m y}$, where $q_m=\frac{2 \pi m }{L}$, and $f_m^n$ are the eigenfunctions of the following problem:

\begin{eqnarray}
	[q_m^4 + \partial_x^4 - 2q_m^2 \partial_x^2] \; f_m^n (x) = \lambda_{m,n}^2 \; f_m^n (x).
	\label{eingenfc}
\end{eqnarray}

This situation is quite similar to the one analyzed in Ref.\ \onlinecite{grapheneacoustic} to solve the classical dynamics for the out-of-plane normal modes. We define the dimensionless variables $\bar{q}_m=W q_m$, $\bar{\lambda}=\lambda W^2$, $\bar{x}=\frac{x}{W}$ and $\bar{y}=\frac{y}{W}$. The solutions of Eq.\ (\ref{eingenfc}) can be written as:

\begin{eqnarray}
  f_m^n (\bar{x}) = \sum_{i=1}^4 d_i \; e^{\beta_i \bar{x}},
\label{fmnd}
\end{eqnarray}

with

\begin{eqnarray}
  \beta_i (m,n) = \pm \sqrt{\bar{q}_m^2 \pm \bar{\lambda}_{m,n}}.
\label{beta}
\end{eqnarray}

After replacing Eq.\ (\ref{fmnd}) in the boundary conditions (\ref{fixbound}) or (\ref{freeboundsig0}) we obtain a linear 4x4 problem whose solutions give $\bar{\lambda}$ as a function of $\bar{q}$, and the coefficients $d_i$ to construct the normalized eigenfunctions. We will elaborate this route in the next section.

Once we have solved this problem, the expression for $h(x,y)$ can be developed as:

\begin{equation}
	h(\bar{x},\bar{y})= \sum_{m,n} \alpha_{m,n} \; f_{m}^n(\bar{x}) \; e^{i \bar{q}_m\bar{y}},
\end{equation}

and with the corresponding change of variables, the path integral (\ref{partfunc}) will run now on the coefficients $\alpha_{m,n}$. The height-height correlation function  $\langle \bar{h}(\bar{x}_1 , \bar{y}_1 ) \bar{h}(\bar{x}_2 , \bar{y}_2)  \rangle$ can be obtained as usual by adding a source term to the exponent of the form $\sum \alpha_{m,n} \epsilon_{m,n}$ and then taking the derivative with respect to $\epsilon_{m,n}$. The result is:

\begin{eqnarray}
\langle \bar{h}(\bar{x}_1 , \bar{y}_1 ) \bar{h}(\bar{x}_2 , \bar{y}_2)  \rangle &=& \frac{kT}{\kappa  \bar{L}} \sum_{m,n}  e^{i(\bar{y}_1 - \bar{y}_2)\bar{q}_m}
\nonumber\\  && \frac{f_m^n (\bar{x}_1) f_{m}^{n} (\bar{x}_2)}{\bar{\lambda}_{m,n}^2}
\label{corrhh}
\end{eqnarray}

with $\bar{h}(\bar{x} , \bar{y} )\equiv \frac{h(W\bar{x} , W\bar{y} )}{W}$. In the previous equation we assumed that the eigenfunctions are normalized in such a way that $\int_{-\frac{1}{2}}^{\frac{1}{2}} \mid f_m^n (\bar{x})\mid^2 d\bar{x}=1$.

\section{Correlation functions for clamped ribbons}
Imposing the conditions of Eq.\ (\ref{fixbound}) over $\bar{h}(\bar{x},\bar{y})$ expressed in terms of $f_{m}^n(\bar{x})$ according to Eq.\ (\ref{fmnd}), we obtain a 4x4 matrix $\mathcal{M}(\bar{\lambda},\bar{q})$ multiplying the vector of coefficients $(d_1,d_2,d_3,d_4)$, whose result must be zero. Explicitly:

\begin{eqnarray}
\mathcal{M} = \left( \begin{array}{cccc}
 e^{\frac{a}{2}} & e^{-\frac{a}{2}} & \cos\left[\frac{b}{2}\right] & \sin\left[\frac{b}{2}\right] \\
 e^{-\frac{a}{2}} & e^{\frac{a}{2}} & \cos\left[\frac{b}{2}\right] & -\sin\left[\frac{b}{2}\right] \\
 a \; e^{\frac{a}{2}} & -a \; e^{-\frac{a}{2}} & -b \; \sin\left[\frac{b}{2}\right] & b \; \cos\left[\frac{b}{2}\right] \\
 -a \; e^{-\frac{a}{2}} & a \; e^{\frac{a}{2}} & -b \; \sin\left[\frac{b}{2}\right] & -b \; \cos\left[\frac{b}{2}\right]
\end{array}\right),
\nonumber\\
\label{sysfixbound}
\end{eqnarray}

where $a=\sqrt{\bar{q}_m^2 + \bar{\lambda}_{m,n}}$ and $b=\sqrt{\bar{\lambda}_{m,n} - \bar{q}_m^2}$ are real variables. By requiring that $\mbox{Det}[\mathcal{M}]=0$ we obtain the values of $\bar{\lambda}_{m,n}$ for each $\bar{q}_m$. The index $n=0,1,2,3...$ used so far enumerates each one of the dispersion curves, which are shown in Fig.\ \ref{reldispfixed}.

\begin{figure}[h]
\includegraphics[trim = 0cm 0cm 0cm 0.05cm, clip, width=20pc]{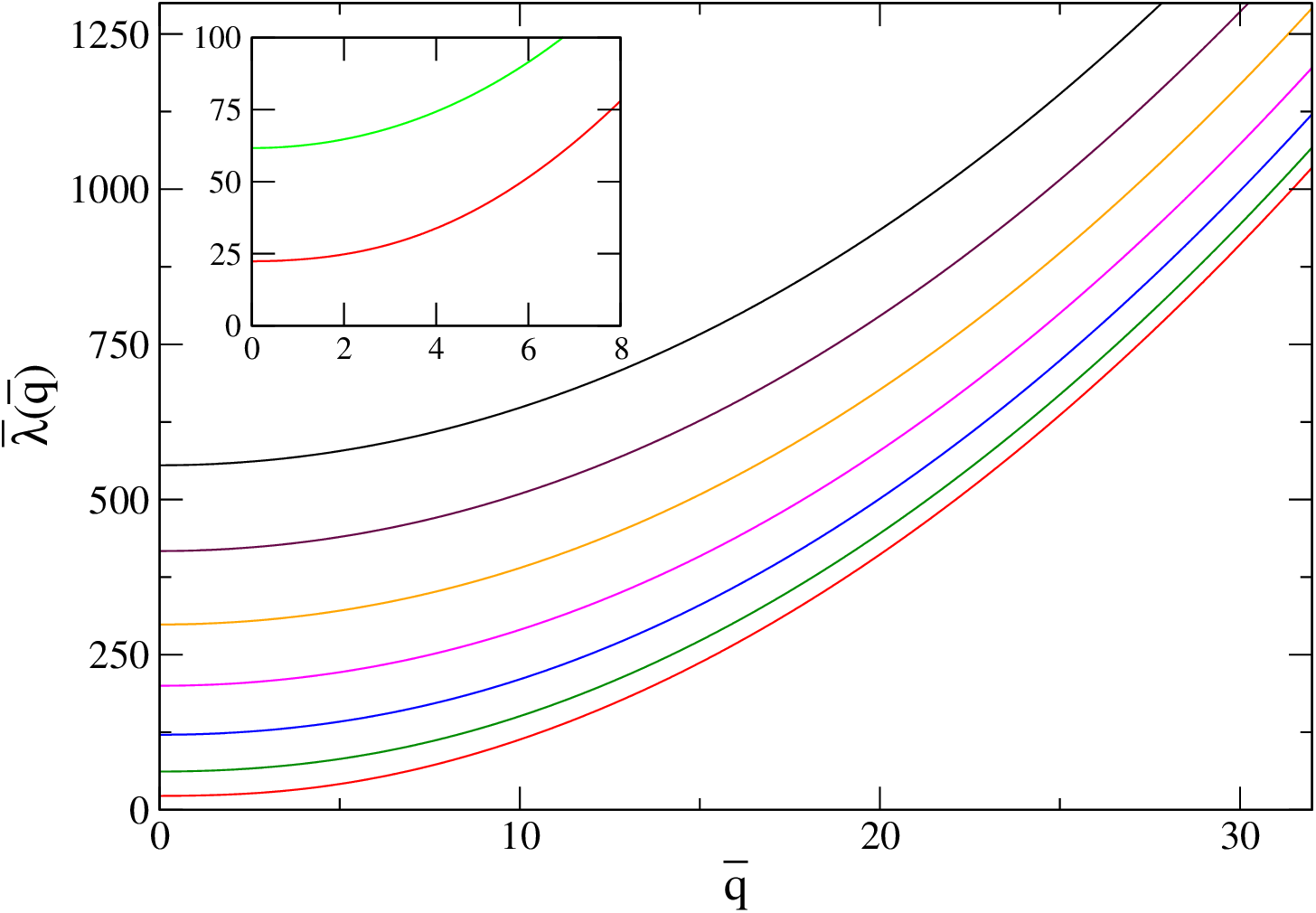}
\caption{\label{reldispfixed} Dispersion curves given by the functions $\bar{\lambda}(\bar{q})$ for the clamped ribbon. We show the first seven branches of the spectrum which, in fact, has an infinite number of them. In the inset we show a zoom of the low energy spectrum for the first two branches.
}\label{reldispfixed}
\end{figure}

These boundary conditions allow us to simplify the determinant of the matrix $\mathcal{M}$, which can be factorized in two terms:

\begin{eqnarray}
\mbox{Det}[\mathcal{M}]&=& -8 \, (a \, \cosh[\frac{a}{2}]\sin[\frac{b}{2}] -b \, \cos[\frac{b}{2}]\sinh[\frac{a}{2}]) \nonumber \\
 && (b \, \cosh[\frac{a}{2}]\sin[\frac{b}{2}]+a \, \cos[\frac{b}{2}]\sinh[\frac{a}{2}]).
\end{eqnarray}

Under these circumstances the matrix $\mathcal{M}$ is reduced to a simpler form and the eigenfunctions $f_{m}^n(\bar{x})$ can be explicitly calculated. When the first term vanishes, we obtain the  $n=1,3,5...$ branches of the spectrum with the odd eigenfunctions, which read:

\begin{eqnarray}
f_m^n (\bar{x}) =C_n \, \left( -\frac{\cos[b/2]}{\cosh[a/2]} \cosh[a\bar{x}]+\cos[b\bar{x}] \right),
\end{eqnarray}

and when the second term becomes zero we obtain the $n=0,2,4...$ ones with the even eigenfunctions, which can be written as:

\begin{equation}
f_m^n (\bar{x}) =C_n \, \left( -\frac{\sin[b/2]}{\sinh[a/2]} \sinh[a\bar{x}]+\sin[b\bar{x}] \right).
\end{equation}

The quantities $C_n$, as remarked at the end of section \ref{GFCF}, represent normalization constants. The plots for $(f_m^n (\bar{x}))^2$ with $n=0,1,2$ and $\bar{q}_m=6\pi$ are shown in Fig.\ \ref{eigenfunclamp}.

\begin{figure}[h]
\includegraphics[trim = 0cm 0cm 0cm 0.05cm, clip, width=20pc]{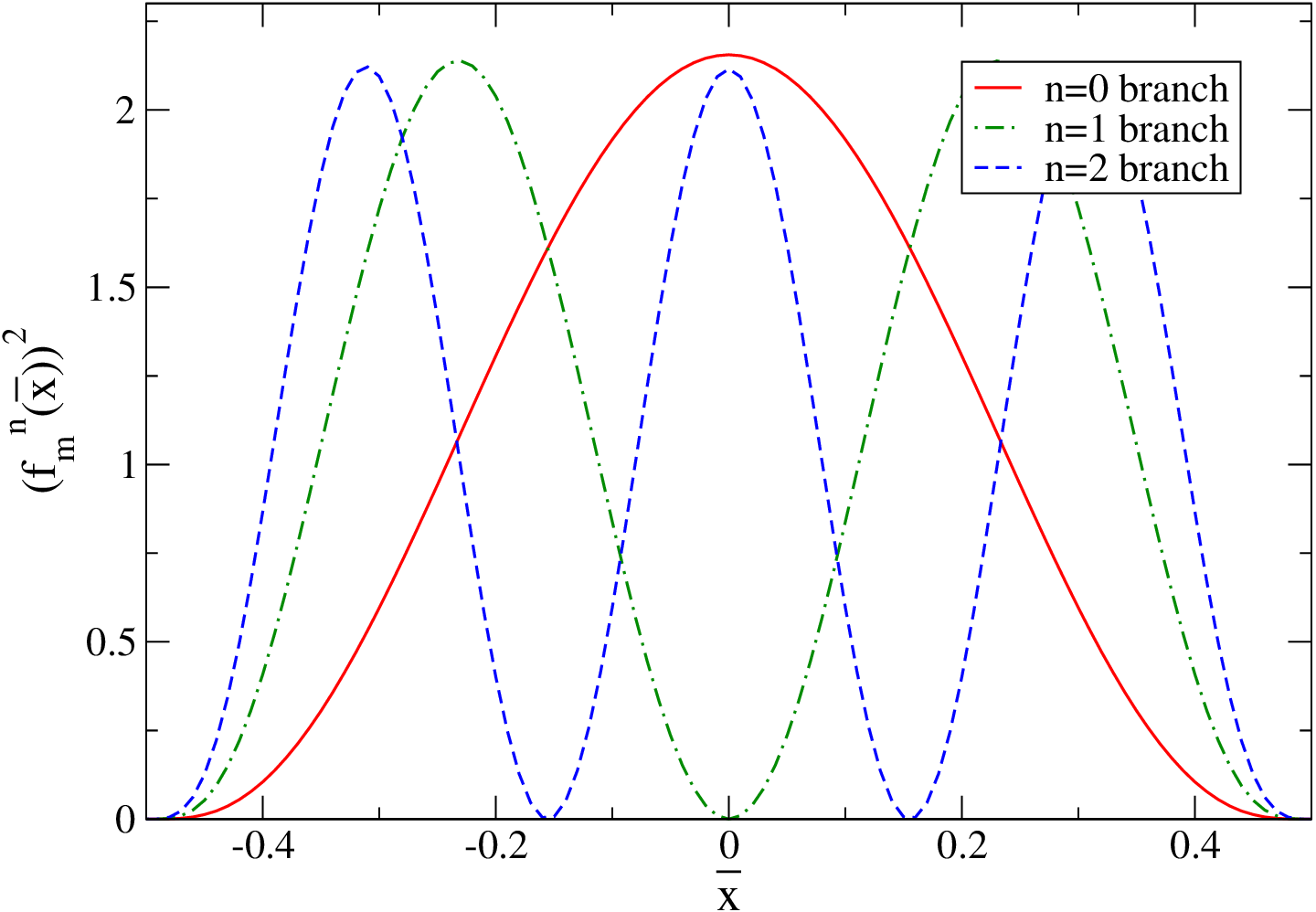}
\caption{\label{eigenfunclamp} Square of the normalized eigenfunctions $f_m^n (\bar{x})$ for the first three branches of the spectrum in the clamped ribbon. The calculations are made for $\bar{q}= 6 \pi$.}\label{eigenfunclamp}
\end{figure}

As has been remarked in Ref.\ {\onlinecite{grapheneacoustic}, there is a gap in the spectrum and the zero energy mode does not exist for $\bar{q}_m=0$. This is related to the fact that global translations are not allowed because the ribbon is gripped by the edges. The gap in the first branch behaves as $\Delta\sim\frac{22.3}{W^2}$ (in the original units) approaching the zero value for the infinite square sheet. We expect that height-height correlations at different points will decay exponentially and this is indeed the case. In Fig.\ \ref{hhfixed} we show the value of $\frac{\kappa  \bar{L}}{k T} \langle \bar{h}(0.25, \bar{y} ) \bar{h}(0.25,0)  \rangle$ running along the $y$ direction and evaluated numerically from Eq.\ (\ref{corrhh}). The contributions of the first three branches are shown. As the gap increases as long as we go to branches with higher energy, the contributions of the corresponding correlations become increasingly smaller.

\begin{figure}[ht]
\includegraphics[trim = 0cm 0cm 0cm 0.05cm, clip, width=20pc]{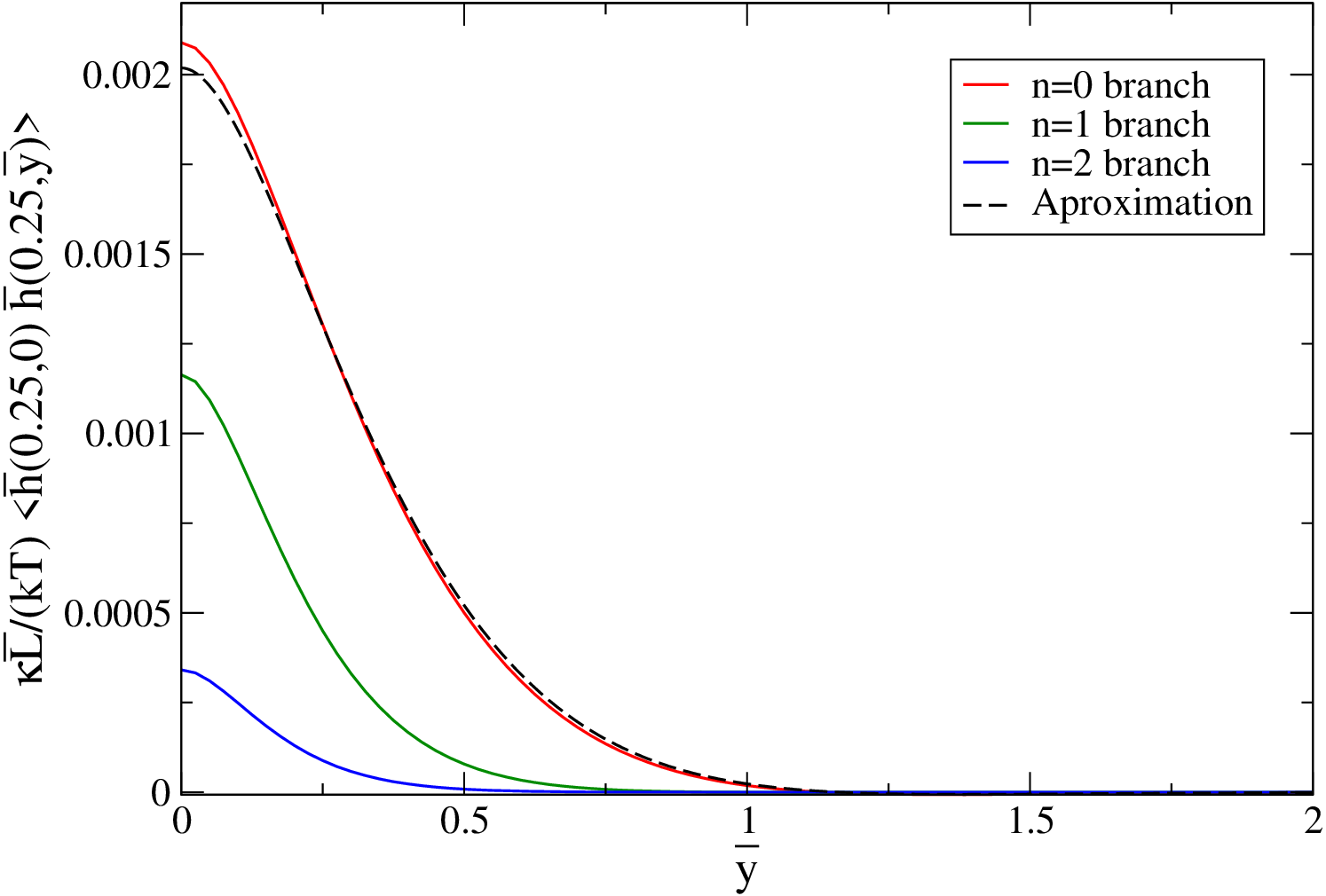}
\caption{\label{hhfixed} Height-height $\frac{\kappa  \bar{L}}{k T}\left\langle  \bar{h}(0.25,\bar{y})\bar{h}(0.25,0) \right\rangle$ correlation as a function of the distance in the long direction, for the clamped ribbon. The contributions of the three first branches are shown separately. The dashed line represents the approximation given by Eq.\ (\ref{hhapp}).}
\end{figure}

We see a rapid decay of the correlations in a distance of the order of $W$. In fact we can estimate the characteristic correlation length with the following approximations. The first branch of Fig.\ \ref{reldispfixed} can be fitted well by a function of the form $\bar{\lambda}_{0}(\bar{q})\simeq \sqrt{a_0+a_1\bar{q}^2+a_2\bar{q}^4}$ with $a_0=500$, $a_1=24$ and $a_2=0.972$. If we neglect the weak dependency  of the eigenfunctions on $\bar{q}_m$ in Eq.\ (\ref{corrhh}), the $y$ dependency of the correlation is given by the following Fourier transform:

\begin{eqnarray}
\langle \bar{h}(\bar{x}_1 , \bar{y} ) \bar{h}(\bar{x}_2 ,0)  \rangle &\simeq & f_m^0(\bar{x}_1) f_m^0(\bar{x}_2) \nonumber \\
&& \int_{-\infty}^{+\infty}\frac{d\bar{q}}{2 \pi}\frac{ e^{i\bar{q} \bar{y}}}{a_0+a_1\bar{q}^2+a_2\bar{q}^4},
\label{fourier}
\end{eqnarray}

which can be analytically solved, giving:

\begin{eqnarray}
\langle \bar{h}(\bar{x}_1 , \bar{y} ) \bar{h}(\bar{x}_2 ,0)  \rangle &\simeq & f_m^0(\bar{x}_1) f_m^0(\bar{x}_2) e^{-q_I \bar{y}} \nonumber \\
&& \left(\alpha \sin \left(q_R \bar{y}\right)+\beta \cos \left( q_R \bar{y} \right)\right),
\label{hhapp}
\end{eqnarray}

where $\alpha=0.00499$, $\beta=0.00271$ and $q_R+iq_I = 2.273+i4.185$ is a zero of the denominator of Eq.\ (\ref{fourier}). The decay of the correlation is clearly dominated by the exponential term. Its characteristic scale i.e. the correlation length, is $\xi=W/4.185$ (in the original units).

We see that it is possible to control the extension of the height-height correlation by changing the width of the ribbon. If we associate this thermal fluctuation with the rippling, this results imply that the characteristic size of the rippled region grows linearly with the width of the ribbons.Let us discuss how the rippling  changes as we move in the $x$-direction. We show in Fig.\ \ref{h2fixed}  the value of $\langle \bar{h}^2(\bar{x},\bar{y}) \rangle $.

\begin{figure}[ht]
\includegraphics[trim = 0cm 0cm 0cm 0.05cm, clip, width=20pc]{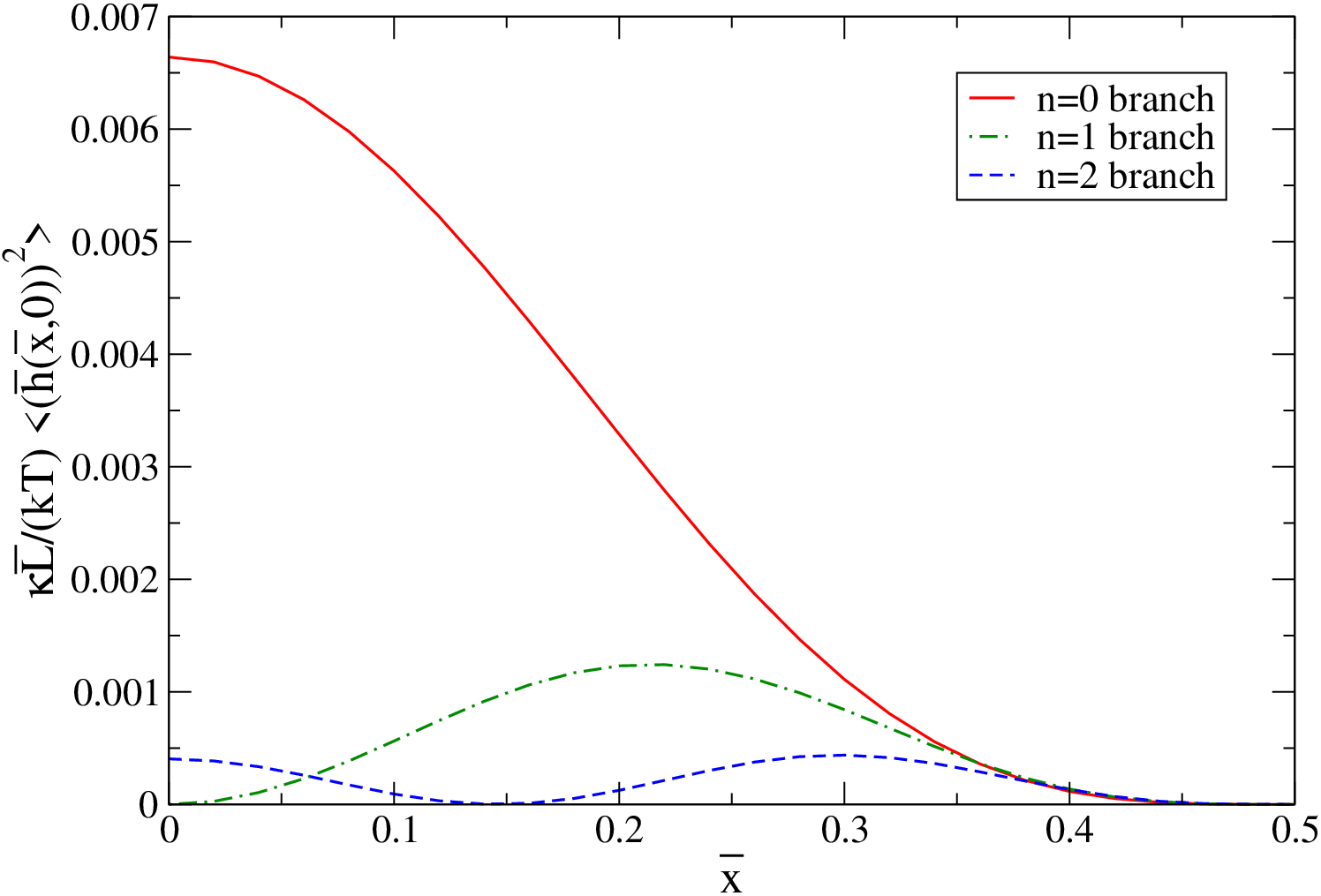}
\caption{\label{h2fixed}  Mean square of the height $\frac{\kappa  \bar{L}}{k T} \langle \bar{h}(\bar{x},\bar{y})^2 \rangle $ as a function on $\bar{x}$, the distance to the center, for the clamped ribbon. We show the contributions of the first three branches. }
\end{figure}

The dominant contribution coming from the first branch produces a maximum distortion at the center of the ribbons. The other branches produce periodic distortions according to the shape of the eigenfunctions $f_m^n(\bar{x})$, as shown in Fig.\ \ref{eigenfunclamp}. The number of  nodes is exactly $n+2$ including those at the edges.

\section{Correlation functions and stability of free edge ribbons}
In a recent paper \cite{CostaDobry} one of the authors has studied the thermal fluctuations and the stability of a graphene rectangle. Periodic boundary conditions were used in both directions in order to avoid edge effects. It has been found, from both analytical calculations and  Montecarlo simulations, that there is a critical relation between the width and the length of the system (called $R_{2D-1D}$) in which the dependence of $\langle \bar{h}^2(\bar{x},\bar{y}) \rangle $ with the system size changes. When the width decreases and the relation with the length becomes smaller than $R_{2D-1D}$, thermal oscillations are more pronounced than in the opposite case. The system behaves as one-dimensional and there is a higher tendency to a crumpled instability  than in square samples.

Moreover, the results of Ref.\ \onlinecite{CostaDobry} do not take into account possible excitations of the edges. To account for this kind of excitations in a more realistic description, we analyze the out-of-plane modes of a ribbon with free boundary conditions, as  given by Eq.\ (\ref{freeboundsig0}). Following similar steps as in the previous section, but with the corresponding boundary conditions, we obtain the dispersion relations of the out-of-plane modes which are shown in Fig.\ \ref{reldispfreesig0}. Differently than in the clamped ribbon, as translational invariance is not broken,  there is no gap but two acoustic branches with zero energy at $\bar{q}=0$.

\begin{figure}[ht]
\includegraphics[trim = 0cm 0cm 0cm 0.05cm, clip, width=20pc]{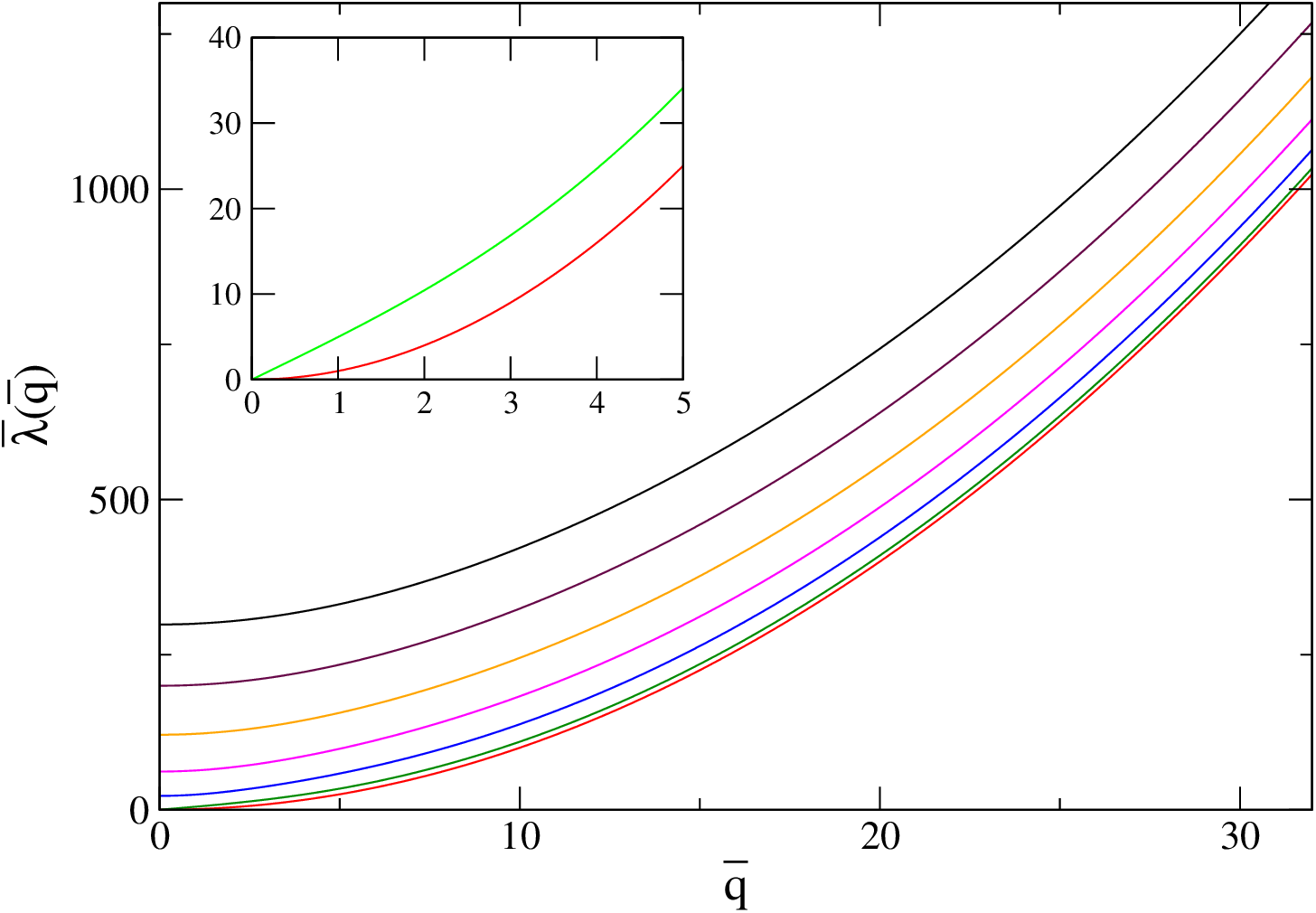}
\caption{\label{reldispfreesig0} Dispersion relations giving the functions $\bar{\lambda}(\bar{q})$ for the free ribbon. We show the first seven branches of the spectrum. In the inset we show a zoom of the low energy spectrum for the two first branches.}
\end{figure}

Qualitatively, we find a similar spectrum as in Ref.\ \onlinecite{grapheneacoustic} where a finite Poisson ratio is taken into account and the boundary conditions given in Eq.\ (\ref{freeboundsigne0}) are used. There are, however, some important differences which appear with a careful inspection of the  two lowest energy branches and their corresponding eigenfunctions. First of all, note that a function of the form $h(x,y)=C e^{iqy}$, in which $C$ is a constant, fulfills condition (\ref{freeboundsig0}) but not (\ref{freeboundsigne0}). This is a solution of the eigenproblem	(\ref{eingenfc}) with a constant eigenfunction and dispersion relation $\bar{\lambda}(\bar{q})=\bar{q}^2$, and this is precisely the eigenfunction corresponding to the lowest energy branch of a system with periodic boundary conditions in the $x$ direction as analyzed in Ref.\ \onlinecite{CostaDobry}. We identify this as a bulk mode. It corresponds to the first branch in Fig.\ \ref{reldispfreesig0}.

This is different than the situation we find when the boundary conditions given in Eq.\ (\ref{freeboundsigne0}) are used. For comparison, we show the square of the eigenfunction corresponding  to the first branch in the upper panel of Fig.\ \ref{eigenfc2branch} for different values of the Poisson ratio $\sigma$. We see that it is a constant function corresponding to a pure bulk mode only for $\sigma=0$, when the model correspond to a membrane. For finite values of $\sigma$ the eigenfunctions depend on the position over the ribbons and do not represent a bulk mode anymore.

\begin{figure}[ht]
\includegraphics[trim = 0cm 0cm 0cm 0.05cm, clip, width=20pc]{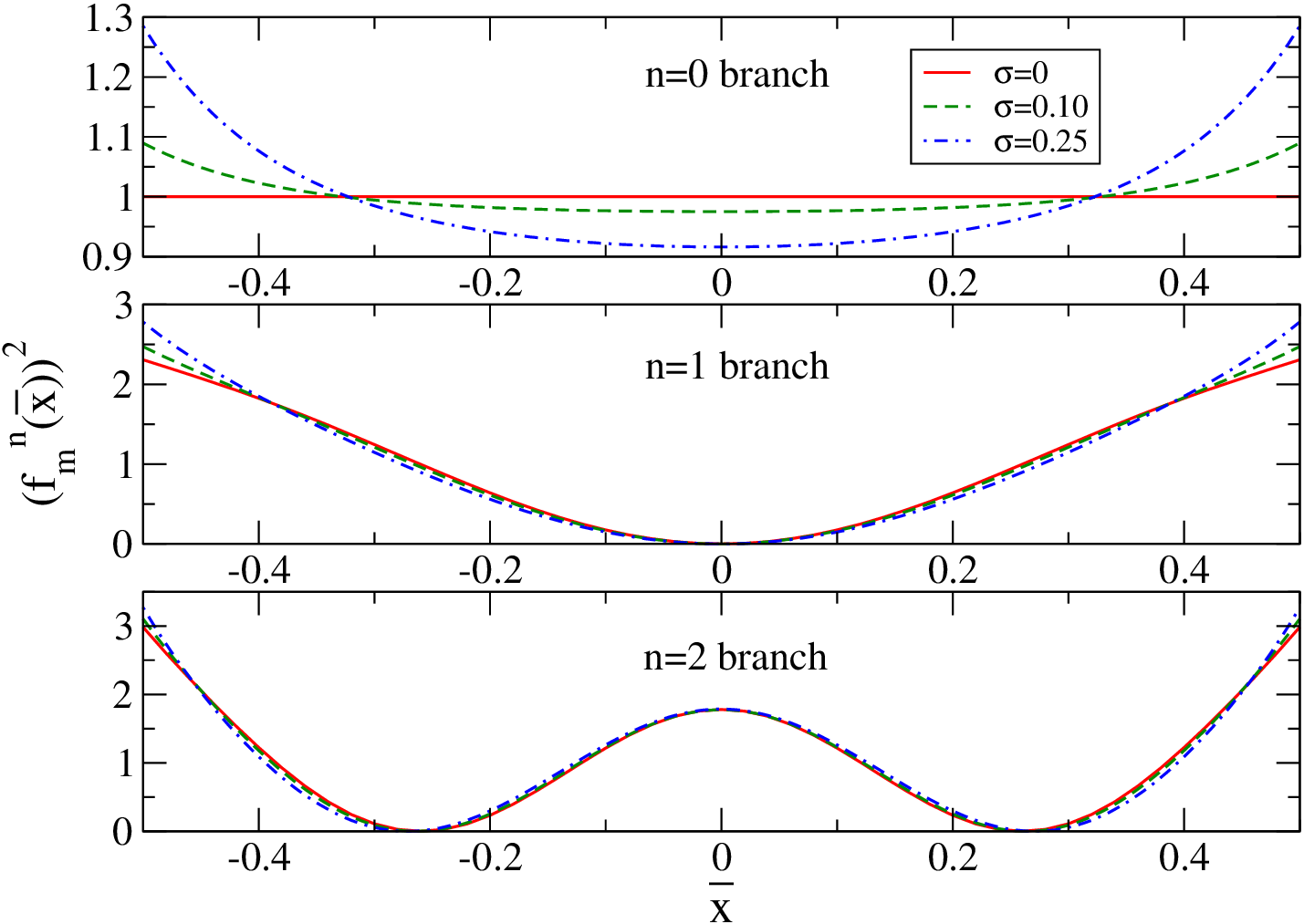}
\caption{\label{eigenfc2branch} Square of the normalized eigenfunctions $f_n^m(\bar{x})$ corresponding to the first (upper panel) second (middle panel)  and third branch (lower panel) of Fig.\ \ref{reldispfreesig0}. We also show for comparison the square of eigenfunctions for  ribbons with finite Poisson ratio $\sigma=0.1$ and $\sigma=0.25$. }
\end{figure}

Otherwise the dispersion relation of the second branch could be well fitted by a law $\bar{\lambda}(\bar{q})=\bar{q} \sqrt{a_0+a_1 \bar{q}^2}$ which disperses linearly near $\bar{q}=0$ and then cuadratically for large values of $\bar{q}$ . The square of the eigenfunction corresponding to these modes is shown in the middle panel of Fig.\ \ref{eigenfc2branch}. We see that the most important distortion occurs when we approach the edges. This mode is therefore mostly an edge excitation and its behavior is qualitatively similar for both $\sigma=0$ and $\sigma \neq 0$.

\begin{figure}[ht] 
\includegraphics[trim = 0cm 0cm 0cm 0.05cm, clip, width=20pc]{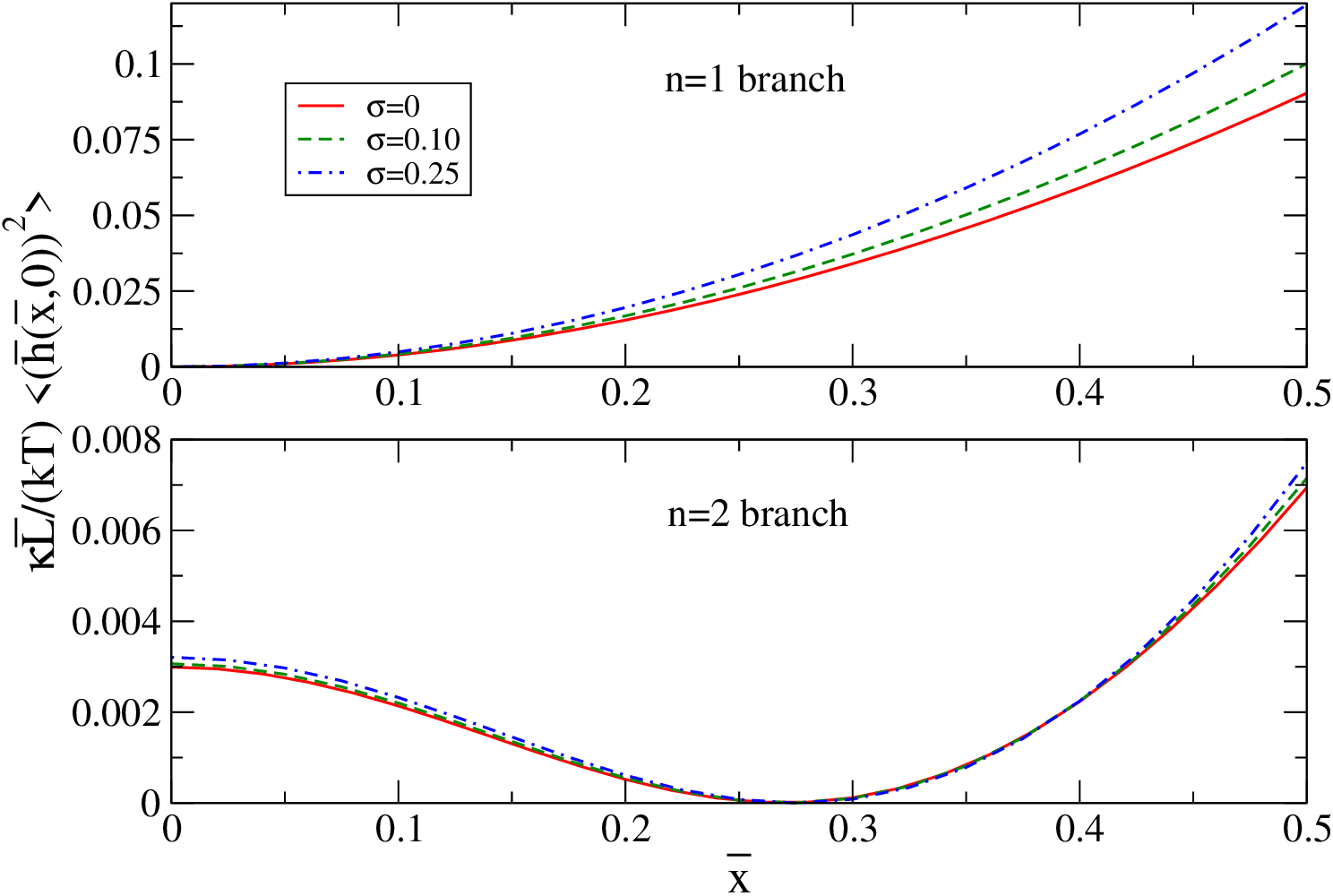}
\caption{\label{h2free}  $\langle \bar{h}^2(\bar{x},\bar{y}) \rangle $ obtained with the eigenfunctions corresponding to the second branch (upper panel) and third branch (lower panel) of Fig.\ \ref{reldispfreesig0}. We also show, for comparison, the corresponding quantities for a ribbon with finite Poisson ratios $\sigma=0.1$ and $\sigma=0.25$.}
\end{figure}

We arrive at the important conclusion: when a theory of a membrane is used, which we assume is the correct one for single- or few-layer graphene systems, the low energy phonon spectrum is decoupled into two branches with quite different physical interpretations. One corresponds to a bulk excitation and the other one is mainly an edge mode.

Let us explore the consequences for the mean square amplitude of the height $ \langle \bar{h}^2(\bar{x},\bar{y}) \rangle $. Differently than in clamped ribbons, there are two gapless branches that give rise to a strong dependency with the length of the system $\sim L$ and a divergence when $L \rightarrow \infty$. This can be interpreted as an intrinsic instability of the system. However, it is known that the harmonic approximation is not  valid for small values of $\bar{q}$ where the anharmonic coupling between the in-plane and the out-of-plane modes\cite{CostaDobry,Los} can stabilize the ribbon (or at least weaken the divergences). Moreover, the external strains which are ubiquitous in real samples, will help to stabilize the system \cite{Roldanstrain}.

The physical effect of the distortion on the ribbon will be quite different depending on the contribution of each of the two lower energy branches. The first one, with quadratic dispersion relation, is a bulk mode. We expect the same effect as the one analyzed in Ref.\ \onlinecite{CostaDobry}, i.e. to produce an homogeneous rippling. More important for our goal in this paper is the effect of the second branch. In Fig.\ \ref{h2free} we show the result for $\langle \bar{h}^2(\bar{x},\bar{y}) \rangle $ calculated using the eigenvectors and eigenvalues corresponding to the second (upper panel) and the third (lowest panel) branch of Fig.\ \ref{reldispfreesig0}. As in the previous figure, we compare this result for ribbons with finite Poisson ratios $\sigma=0.1$ and $\sigma=0.25$ (using  the values corresponding to the second and the third branches of its dispersion relation). We see again that the maximum of $\langle \bar{h}^2(\bar{x},\bar{y}) \rangle $ is at the boundary for each value of $\sigma$. Regarding the contribution of the third branch, the correlation is strongly reduced as there is a gap for these excitations. A minimum around $\bar{x}\sim 0.28$ is observed as well, corresponding to a node in the eigenfunctions.

\begin{figure}[ht]
\includegraphics[trim = 0cm 0cm 0cm 0.05cm, clip, width=20pc]{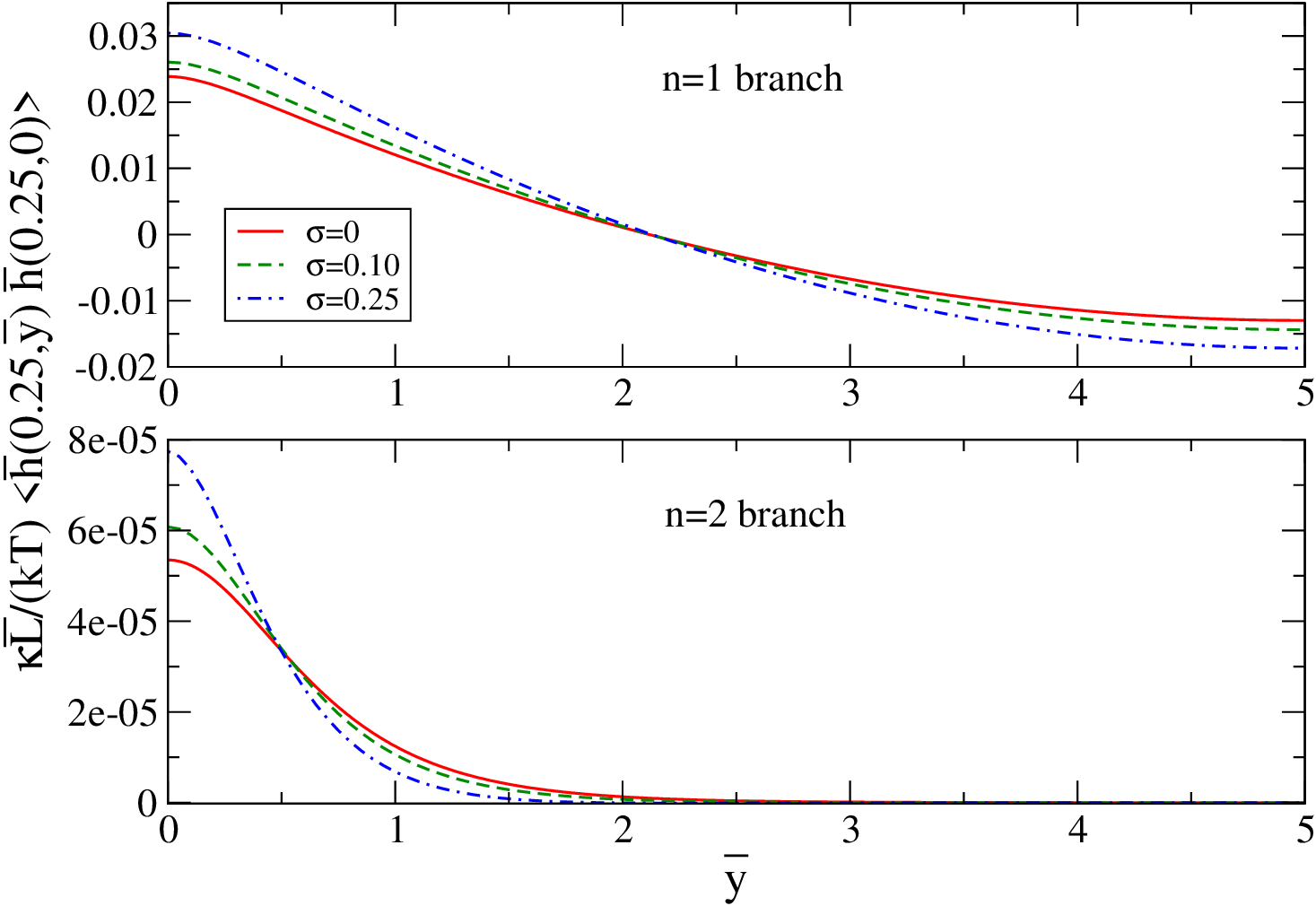}
\caption{\label{hhfree} Correlation $\langle h(0.25,0)h(0.25,y) \rangle $ obtained with the eigenfunctions and eigenvalues corresponding to the  second branch (upper panel) and the third branch (lower panel) of Fig.\ \ref{reldispfreesig0}. We also show, for comparison, the corresponding quantity for ribbons with finite Poisson ratios $\sigma=0.1$ and $\sigma=0.25$. The length of the ribbons is $L=1000$ and its width $W=100$.}
\end{figure}

The height-height correlation function is shown in Fig.\ \ref{hhfree} as a function of $\bar{y}$ for fixed $\bar{x}=0.25$. Regarding the contribution of the third branch in the lower panel, we see a similar behavior than the one seen in the clamped ribbon analyzed in the previous section. The correlation function has very small values and it decays exponentially with the $\bar{y}$ distance. However, the contribution of the second branch is much more important and the correlation function decays much more slowly. This is a consequence of the absence of a gap for this mode. In fact, the ultimate value that this correlation assumes, increases with the length of the ribbon.

Finally, let us  briefly discuss the consequence of this result for a distortion of a free standing sample. It has been seen from  early measurements that when  graphene samples are put in a scaffold configuration, the free edges appear folded\cite{meyer}. The fact that a scrolled configuration can be a stable deformation of a graphene edge has been previously shown\cite{castroscroll}. It is stabilized by an interplay between the van der Waals  and  bending energies. The edge modes found in the present paper could be interpreted as precursive modes for this large distorted structure. That is to say, one possible sequence of events might be: first the edge oscillates rapidly with increasing amplitude and second it sticks to the rest of the sample due to the attractive van der Waals force.

\section{Conclusion and Discussion}
In this paper we have studied the out-of-plane phononic spectrum and the height-height correlation functions of a graphene nanoribbon. Two different configurations were considered: clamped at the edges and free edges. When the ribbon is gripped, there are no true acoustic branches but a gap in the  phononic spectrum. This gap leads to an exponential decay of the correlation and thermal excitations produce only local distortions of the crystalline order. This looks quite similar to the situation in usual three dimensional solids. However, when the lateral dimension $W$ increases enough, the characteristic correlation length also increases. This is coherent with the fact that in a square infinite membrane, the  height-height correlation decays as a power law \cite{Los}.

When the edges are free, we find a quite remarkable decoupling between two different low energy phononic branches. Both of them go to zero energy for $\bar{q} \rightarrow 0$. One disperses quadratically for all values of $\bar{q}$. The eigenvectors corresponding to this branch do not depend on the transversal coordinate of the ribbon. Therefore, in these modes the system is unaware of the existence of the edges. It is in fact a bulk excitation, similar to a lower excitation of an infinite membrane. The possible instability connected with this mode could lead the system to an homogeneous rippling. The other low energy branch disperses linearly for small values of $\bar{q}$. The corresponding eigenvectors have their maximum values on the edges. It is a surface or edge mode. An instability is also associated with these modes. We claim that they could be the precursive modes resulting in folded edges as seen experimentally in suspended graphene samples. We remark that this decoupling between a bulk an edge mode is specific of a membrane-like model. In a model for a continuum plate in which a finite  thickness is assumed, both modes are essentially edge modes. As graphene is a one-atom thick membrane, we expect that this fact would be specially significant for graphene nanorribons.

A possible extension of our work could be the study of the interactions between the phononics modes and the conducting electrons in nanoribbons. It has been shown that flexural phonons play an essential role as a limiting mechanism for the mobility in suspended graphene. We presume similar effects in nanoribbons. Moreover, when the edges of the ribbon are of zig-zag type, localized electronic states near the edge are expected \cite{brey}. The phononic modes found in the present work should interact considerably with these confined electronic states. We expect them to play an important role in the transport properties for these nanoribbons.

\section{Acknowledgments}
We thank D. Mastrogiuseppe for critical reading of the manuscript.
This work was partially
supported by PIP  11220090100392 of CONICET, and  PICT R 1776 of the ANPCyT.

\end{document}